\documentclass[10pt,twocolumn,letterpaper]{article}

\usepackage{cvpr}
\usepackage{times}
\usepackage{epsfig}
\usepackage{graphicx}
\usepackage{amsmath}
\usepackage{amssymb}
\usepackage{subfigure}

\usepackage[breaklinks=true,bookmarks=false]{hyperref}

\cvprfinalcopy 

\def\cvprPaperID{W19-12} 

\begin{document}

\title{Detecting Clues for Skill Levels and Machine Operation Difficulty from Egocentric Vision}
\author{Longfei Chen, Yuichi Nakamura, Kazuaki Kondo \\ 
Academic Center for Computing and Media Studies, Kyoto University\\
{\tt\small yuichi@media.kyoto-u.ac.jp} 
}

\maketitle

\begin{abstract}
With respect to machine operation tasks, the experiences from different skill level operators, especially novices, can provide worthy understanding about the manner in which they perceive the operational environment and formulate knowledge to deal with various operation situations. 
In this study, we describe the operator's behaviors by utilizing the relations among their head, hand, and operation location (hotspot) during the operation. 
A total of 40 experiences associated with a sewing machine operation task performed by amateur operators was recorded via a head-mounted RGB-D camera. 
We examined important features of operational behaviors in different skill level operators and confirmed their correlation to the difficulties of the operation steps.  
The result shows that the pure-gazing behavior is significantly reduced when the operator's skill improved. Moreover, the hand-approaching duration and the frequency of attention movement before operation are strongly correlated to the operational difficulty in such machine operating environments. 
\end{abstract}

\section{Introduction}
Many studies have explored operational behaviors in various environments, such as kitchens and offices \cite{inwhatways, coordination, youdo}. 
The standard or efficient behaviors of operators, particularly from specialists, are good resources for task modeling, providing guidance, skill assessment, etc. 
In addition to those essential behaviors, the behaviors of operators of various skill levels, especially novices, are also good resources that delineate their perception the environment, formulate knowledge, and operation difficulty.
For instance, the operators may search for future-use objects, hesitate, and make mistakes. They may concentrate on the operation location and verify the outcomes \cite{inwhatways}. Information extracted from these actions can be used for task and guidance design.

In this paper, we introduced an automatic analysis for the aforementioned purpose. We chose a sewing task as an example and recorded operation experiences of different skill levels using a head-mounted RGB-D camera. The relations between the attention, hand, and hand--machine interacting area (hotspot) were examined to detect the intrinsic characteristics of the operations. Our preliminary experiments successfully showed close relationships between operation behaviors, operation difficulties, and the operators' skill levels.

\begin{figure}
\begin{center}
\setlength{\belowcaptionskip}{0.5cm}   
\includegraphics[width=0.39\linewidth]{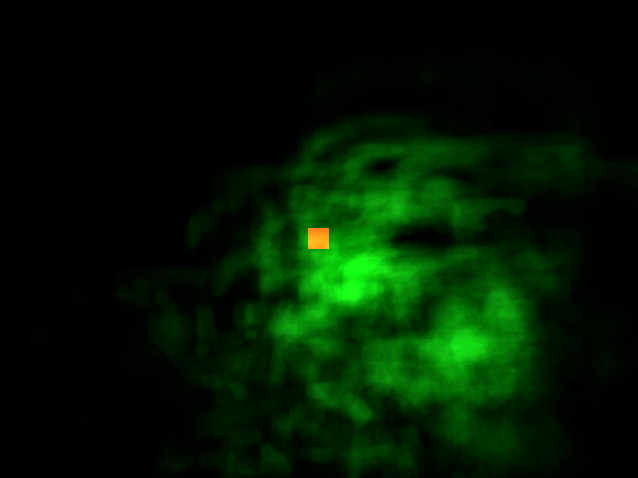}\quad \ \ 
\includegraphics[width=0.39\linewidth]{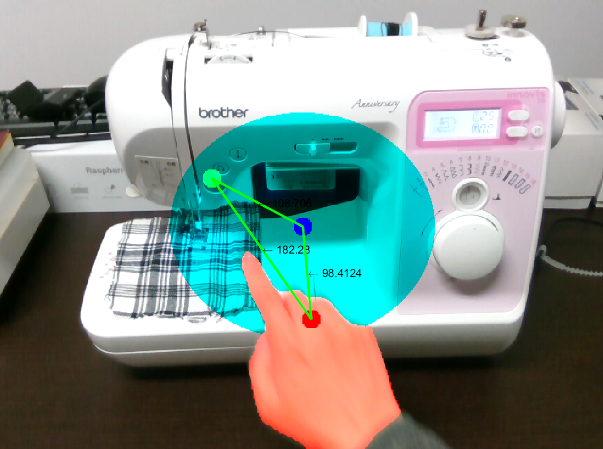}
\end{center}
\caption{[L] The accumulated touches of 40 sewing experiences (\emph{green}) are distributed around the center of FPV camera view (\emph{red}) with a little bias.
[R] The 2D distances between the hand (\emph{red}), view center (\emph{blue}) and the hotspot (\emph{green}) are automatically extracted to describe operation behaviors, indicating operation difficulties and user skills.}
\label{fig:method}
\end{figure}

\section{Related work}
Previous studies have focused on the relation between gaze and operation characteristics. Land and Hayhoe \cite{inwhatways} used an eye tracker in the task of making tea and sandwiches. Pelz et al. \cite{coordination} monitored the eye, hand, and head coordination in a block-copying task. Such tasks could usually be divided into a series of object-related actions (ORAs), and the eye movement primarily leads each action on a top-down control of `as needed' basis.
The actions manifest into a temporary synergistic linkage of eye leading, followed by head and then hand. One noteworthy observation is that the gaze is often shifted to the next object in sequence before the completion of the current activity.
We need a reliable and automatic measurement to investigate  the intricate temporal, spatial patterns, and large variations in the operation behaviors \cite{vision}.
Other studies investigated the relation between operation behaviors and skill levels. Zhang et al. reported a video-based skill evaluation in surgical training using motion features \cite{relativehmm}. Hazel et al. proposed a supervised deep ranking model to determine skills in video records in a pairwise manner \cite{whosbetter} using CNNs. However, we still lack reliable semantic explanations of which features have close relations with skill levels and how they related.

\section{Operation behaviors and their features}
\subsection{Characteristics of operations}
In our experimental environment, operators sit in front of a table manipulating the sewing machine with all the needed materials within their reach. This is a quick process without considerably long waiting. 
User's operation behaviors can provide information regarding the task process and its difficulty.\\
\textbf{Operator's skill level:} A novice may require a many searches to formulate the memory of task-relevant objects \cite{inwhatways}, or hesitate to perform the next step. However, a skilled operator can finish the task quickly with a little checking, or shifting their attention to the next target before finishing the current operation.\\
\textbf{Operation difficulty:} An easy operation can be performed without much attention; whereas a difficult operation requires a lot of monitoring, concentration, checking, and confirmation \cite{fivestage}. 

These situations heavily affect user's operation behaviors. 
A target of gaze is one of the most important spots. It is a strong cue of the operator's attention.
The hand is another important spot, which indicate the operators' intentions and procedures to perform operations. 
The third important spot is ``hotspot'', which represents the frequent-interacting location between a hand and a machine \cite{hotspots}. 

Fig.\ref{fig:method} (L) illustrates the accumulated touches, distributed around the center of the FPV camera with a little bias toward the right-bottom\footnote{It was caused by the operators who mostly use the right hand. This denotes a similar center bias regarding eye--hand coordination in \cite{predictgaze}.}.
Touching at the center of view may represent careful and concentrating operation, and touching far from the center may mean operation step with little attention.
Similarly, the duration, correlations, and latencies of user's head and hand moving towards hotspots are also used to analyze their behaviors \cite {inwhatways, coordination}. 
Fig.\ref{fig:method} (R) shows the spatial relations among egocentric view center, hand, and hotspots in an operation.

\subsection{Feature extraction}

\begin{figure*}[t]
\begin{center}
\subfigure[search]{\includegraphics[width=0.24\linewidth]{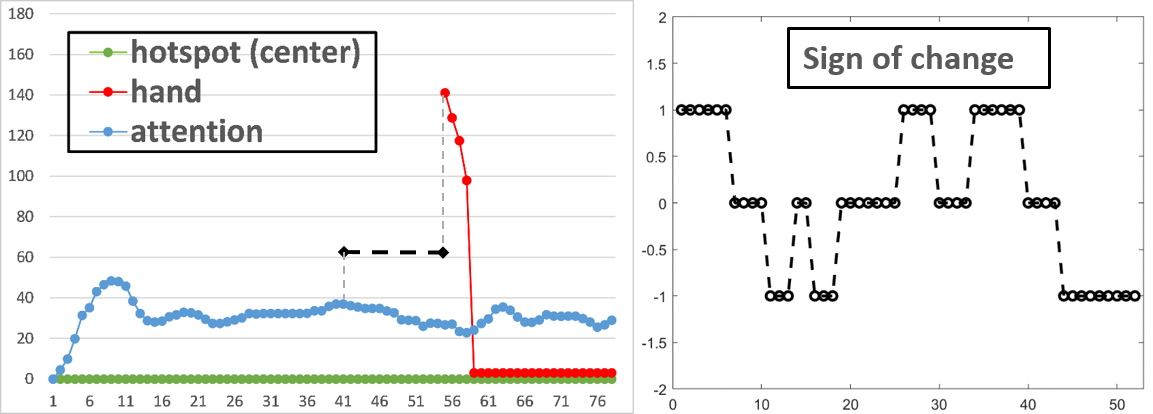}} \ 
\subfigure[shift]{\includegraphics[width=0.24\linewidth]{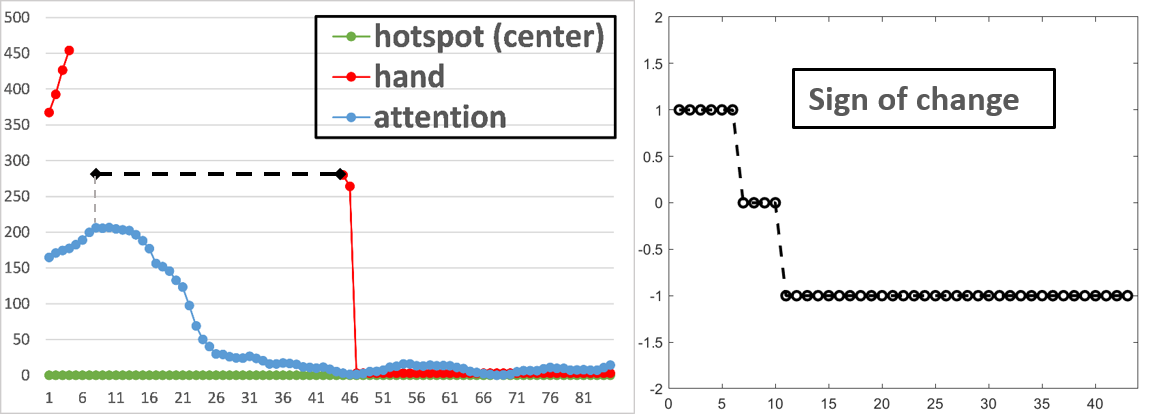}} \ 
\subfigure[early shift]{\includegraphics[width=0.24\linewidth]{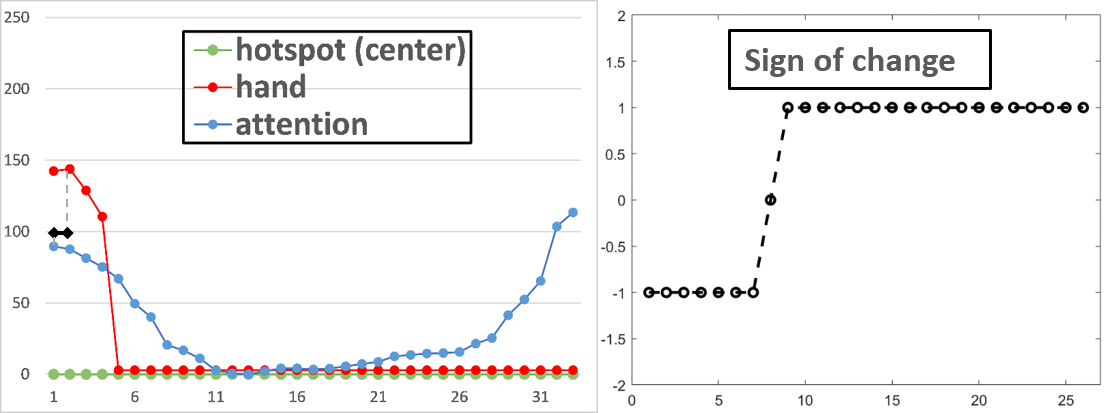}} \ 
\subfigure[non-early shift]{\includegraphics[width=0.24\linewidth]{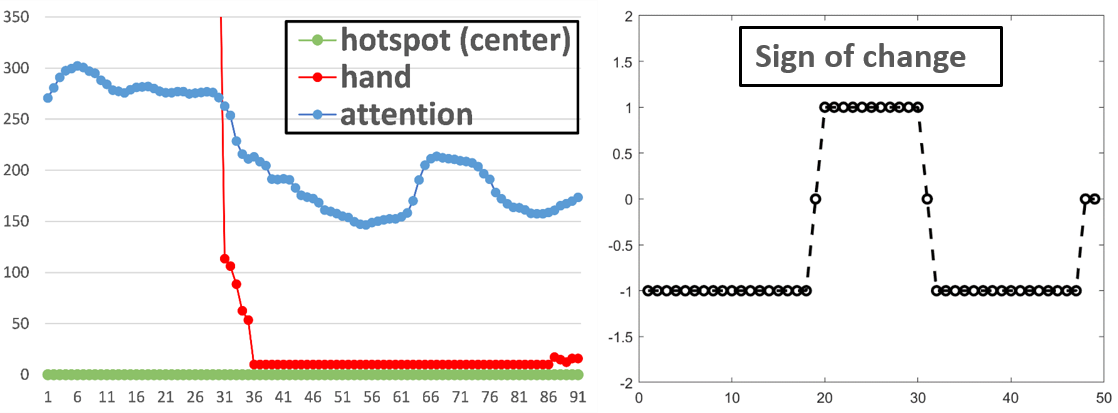}}
\end{center}
\caption{Different patterns of gaze behaviors. Left sub-figures show the distances of estimated attention and hand respect to the hotspot (as coordinate origin). The black-dotted line shows the lag of attention precedes hand that approaches the hotspot. $d_{HO}$ is set as 0 when the hotspot is touched by hand. Right sub-figures denote the sign of distance change. In search (a), the sign of attention--hotspot distance ($d_{AO}$) changes frequently; however, in the case of gaze shift (b), $d_{AO}$ simply decreases. Figure (c) and (d) show early shift and non-early shift behaviors, distinguished by whether a continuous increase of $d_{AO}$ in the late period of operation.}
\label{fig:gazepatterns}
\end{figure*}

\paragraph{Gaze:} Studies \cite{predictgaze, tessid} have reported strong correlations between the gaze and the head movement in such egocentric operation environments. 
The egocentric gaze often aligns with the head orientation that exhibits a considerably small variance in space, and the center of the view can give a rough direction towards the first-person's head orientation. 
Hence, we may simply utilize the center of the FPV camera to direct the operator's attention direction, and use the head movement to approximate the attention shift. 
To adjust the user's initial attention, we lead the operator's gaze with a red point and posit it at the center of the egocentric view by adjusting the head-mounted camera position. Though small offsets may exist between the recording center and actual attention location, we try to compensate the offsets and do not consider the exact attention location, but mainly consider the relative movement of attention to hand and hotspot, which is considered to be useful for describing several gaze behaviors, e.g., \emph{shift}, \emph{search}, or \emph{concentrate}. In addition, the divert of attention orientation, the speed, frequency, and variance of attention shift can further describe the operators' attention patterns. 
\\
\textbf{Hand and hotspots:} To detect a hand, we first segment the foreground by considering a common operation distance; then, a skin color model is built for each user at the initial period of operation to extract the hand region. 
A hotspot is detected by clustering the touches of between the hand and machine surface utilizing both spatial and temporal locations. The details are given by \cite{hotspots}.
\\
\textbf{Temporal duration:} The operation experience is divided into a sequence of operation units (OUs). Each unit is a combination of sequential periods of ``pure-gazing (\emph{saccade/fixation}) $\rightarrow$ hand-approaching $\rightarrow$ operating” \cite{inwhatways}. Although their duration may vary significantly.
The pure-gazing period is considered to be in between the end of the previous physical contact and the moment when the hand goes into the sight. The hand-approaching period can be considered as the period between the moment the hands just appear and the moment at which the operation begins. Subsequently, the operating period is the period in which physical touches occur. For each OU, the absolute duration of each period and their relative ratios to the OU are measured. 
\\
\textbf{Distance:} The 2D distance between the positions of estimated attention (A), hand (H) and operation (O) are calculated. 
For simple denotation, we denote their position as $p_{A}$, $p_{H}$ and $p_{O}$, and the distance between each pair of the locations is denoted as $d_{AO}$, $d_{HO}$ and $d_{AH}$ in following sections. 
To compensate the offsets between the center of the view and the actual attention location for each OU, we use:
{\setlength\abovedisplayskip{3pt}
\setlength\belowdisplayskip{3pt}\begin{equation}
d^* = d - d_{0},
\end{equation}}
where $d^* $ is the estimated distance, $d$ is the measured distance, and the offset $d_{0}$ is the minimum value of $d$ in this operation period.
\\
\textbf{Speed, frequency, variance:} We describe the spatial tendency of two areas, e.g.,  whether they are getting closer or distant, or how fast and frequent the area moves towards each other, with the sign of distance change ($+$, $0$, $-$) and the changing speed, frequency, and variance. These are computed as
{\setlength\abovedisplayskip{3pt}
\setlength\belowdisplayskip{3pt}
\begin{equation}
\begin{aligned}
v = \emph{diff}(d^*), \:\: 
f = \mathbb{C}(\emph{sign}[v]), \:\: 
\text{and} \:\:  
\delta^2 = E[(d^* - \bar{d})^2] .
\end{aligned}
\end{equation}}
where $d^*$ is the estimated distance and $\bar{d}$ is its mean, and $\mathbb{C}$ is the number of sign changes of a period. 
As shown in Fig.\ref{fig:gazepatterns} (a), the operator's head oscillated back and forth when searching for the hotspot of next use, the sign of attention--hotspot distance ($d_{AO}$) is changing frequently; 
while when the operator directly shifts attention to the next, $d_{AO}$ is decreasing monotonously (Fig.\ref{fig:gazepatterns} (b)). 
\\
\textbf{Correlations:} To investigate the synergy of attention and hand during an operation, we compute the correlation coefficient of their distances to the hotspot ($d_{AO}$ and $d_{HO}$). 
The temporal latency of attention proceeding hand is also calculated base on the timing difference when approaching the hotspot.
\\
\textbf{Early shift:} The gaze is often observed to shift to the next operation location before the completion of the current operation process in an average of $0.61s$ for the tea-making task \cite{inwhatways} or $100ms$ earlier for a relatively skilled player in a cricket game \cite{play}. 
Therefore, we consider that the ratio of the early-shift of an operation is an important indicator of the operator's familiarity with this operation. 
We calculate the early-shift ratio of an operation by considering the increase of the attention--hotspot distance ($d_{AO}$) at the later period during the operation:
{\setlength\abovedisplayskip{3pt}
\setlength\belowdisplayskip{3pt}\begin{equation}
R = p_{1}/p_{o}, \:\: s.t.\:\:  p_{o}\geqslant1s
\end{equation}}
where $p_{1}$ is the length of continuous ``$+$'' sign of $d_{AO}$ at the finishing phase of an operating period, and $p_{o}$ is the total duration of the period. 
As shown in Fig.\ref{fig:gazepatterns} (c), the location of the attention approached the hotspot at the beginning of the operating period; however, at the finishing phase of operation, it shifted away from the hotspot, which manifests a continuous increase of $d_{AO}$. 

\section{Experiment}
We gathered 20 pairs of egocentric operation experiences from 16 beginners of a sewing task using a head-mounted RGB-D camera. 
Each pair comprises an earlier experience and a later experience of the same operator. 
We consider the later experiences to be overall more skilled than earlier experiences despite the single pair variations. 
The aforementioned features were extracted for each experience. Operating periods less than $1s$ are ignored for detecting early-shift in our experiment. 
We compared the features in experiences with different skill levels, and investigated their correlations with the subjective evaluations of operational difficulties. 
To obtain the subjective evaluation of the operation difficulties, three experts and three beginners are asked to rate the difficulty score of each operation step from $-$5 to 5 (most difficult to easiest).

\subsection{Comparison at different skill levels}
\begin{figure*}[ht]
\begin{center}
\subfigure[]{\includegraphics[width=0.32\linewidth]{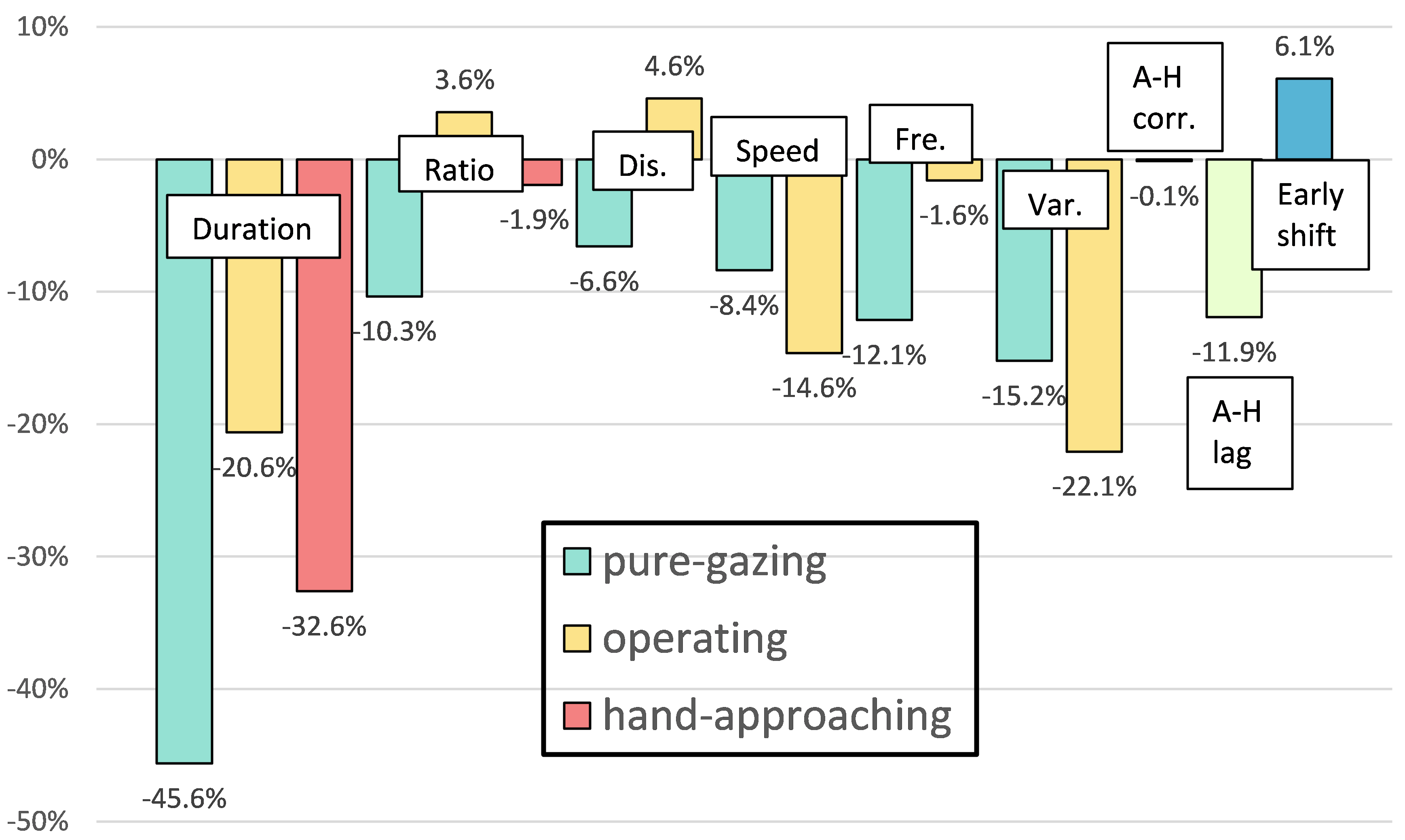}} \ \ \ \ \ \ \ \ \ \ 
\subfigure[]{\includegraphics[width=0.32\linewidth]{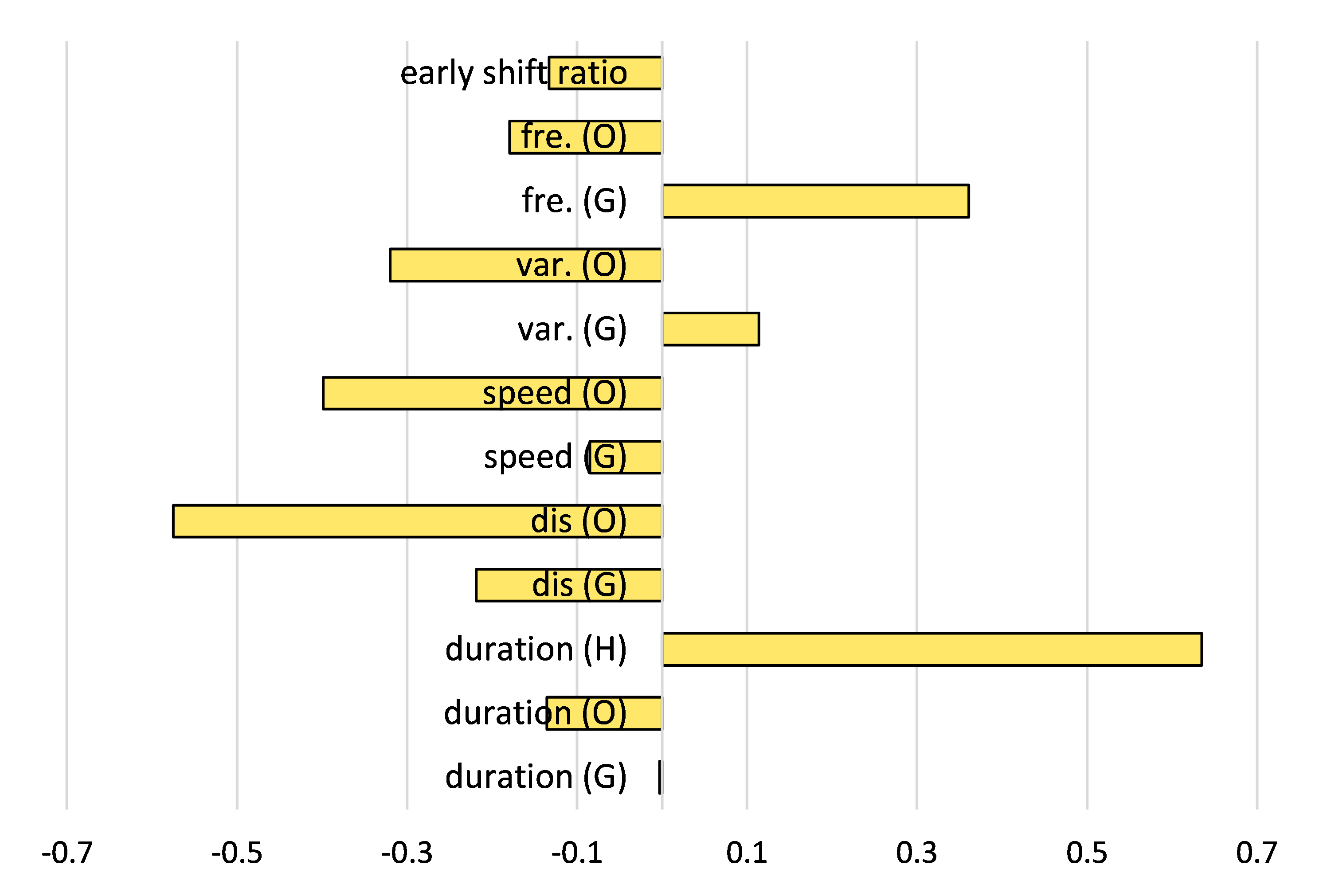}}
\end{center}
\caption{(a) Comparison of features between earlier and later experiences (averaged with 20 pairs). The average of all earlier experiences are set as the baseline (0\%), and the relative ratios of later experiences are shown for comparison. (b) The correlation of features to user-rated operation difficulties, where G, H, and O denote the pure-gazing, hand approaching, and operating periods, respectively.}
\label{fig:correlation}
\end{figure*}

To reduce inter-person variation, we compared the operation features of different skill level experiences in a pair-wise manner, i.e., we compare the ratio of a later experience relative to an earlier experience in each pair, then average for all pairs. Fig.\ref{fig:correlation} (a) shows the comparison result. 
For simple denotation, we refer to the less skilled beginners as “Novice” and the more skilled beginners as “Advanced beginner” by referring to the Dreyfus model \cite{fivestage}. 

The absolute duration of pure-gazing, hand-approaching and operating periods in later experiences are considerably reduced by more than 45\%, 32\%, and 20\%, respectively, 
which indicates the advanced beginner completed the task much quicker with increasing familiarity. 
Considering the relative ratio of each period to the OUs, the pure-gazing is still the most significant difference when skill improved, which reduced 10.3\%; 
while the operating and hand-approaching ratio varied slightly ($+$3.6\%, $-$1.9\%).

We examined the factor responsible for a reduction in pure-gazing in more skilled experiences. The frequency of attention movement is significantly reduced ($-$12.1\%) as compared with the operating period ($-$1.6\%). This result is consistent with the assumption that a novice requires more searches to build the memory of relevant information before the operation.

During the operating period, 14 out of 20 earlier experiences demonstrate average smaller attention--hotspot distance, indicating that the novice concentrate more on the operating spot. 
However, their attentions are fiercer regarding the higher speed and variance, probably caused by the result-checking behaviors.

No difference in the correlation of $d_{AO}$ and $d_{HO}$ between novices and advanced beginners, indicating that the spatial synergy of hand--attention remains similar in different skilled operators. 
The time lag of user's attention precedes hand to reach the hotspot reduced from averagely $0.21s$ to $0.18s$ from earlier to later experiences, which showed a smaller attention--hand latency of quicker action implementation when skill improved.

With respect to the early-shift ratio, later experience shows relatively larger values ($+$6.1\%), revealing that advanced beginners shift their gaze earlier to next steps before finishing the current operations.

\subsection{Operation difficulty}

We then examined if the features indicate the difficulty of operation. 
Fig.\ref{fig:correlation} (b) depicts the correlation coefficients of the features and the rated difficulty scores for the 15 essential operation steps of the sewing task. 

The result show that (i) the duration of hand-approaching period and (ii) the frequency of attention movement in pure-gazing period are strongly correlated to operation difficulty. 
This implies that the more difficult an operation step is, the more frequent the operator searches before operation, and the longer operator's hand approaching the operation location. 
This is probably because that the hand-approaching time is closely related to the hesitation of an operator in this quick-tempo and small-space operation task; however, sacaade for search is crucially required for unskilled operators.

On the other hand, the high-negative correlated features are (i) larger attention--hotspot distance in operating periods, and (ii) high speed of attention movement during operation. This implies that the less concentration, the faster the gaze shift, the easier the operation is for the operator. 
However, the early-shift behavior only shows a weak negative correlation with the difficulty in our experimental results.

\section{Conclusion}
In this paper, we introduced our approach for automatically detecting characteristics of operations and operators. 
We picked several features derived from the relationships of  head, hand, and hotspots, and confirmed if they are good indexes of the skill level of the operators and  operational difficulties. 
The result of our preliminary experiments shows good potential of those features. 
For future work, systematic data collection and verification are necessary to clarify their characteristics. 
Further, the supporting task modeling or task guidance are also attractive topics.

{\small
\bibliographystyle{ieee_fullname}
\bibliography{egbib}
}

\end{document}